\documentclass[lettersize,journal]{IEEEtran}
\usepackage{amsmath,amsfonts}
\usepackage{algorithmic}
\usepackage{algorithm}
\usepackage{array}
\usepackage[caption=false,font=normalsize,labelfont=sf,textfont=sf]{subfig}
\usepackage{textcomp}
\usepackage{stfloats}
\usepackage{url}
\usepackage{soul}
\usepackage{verbatim}
\usepackage{graphicx}
\usepackage{multirow}
\usepackage{epstopdf}
\usepackage{cite} 
\usepackage{flushend}
\usepackage{xcolor}
\usepackage{flushend}
\usepackage{hyperref}
\hypersetup{
    colorlinks=true,
    linkcolor=red,
    filecolor=blue,      
    urlcolor=blue,
    citecolor=green,
}

\begin{document}

\title{Flexible Deep Joint Source-Channel Coding: A Vibrotactile Example}

\author{Shuijie Li, Kemi Chen, Runjie Wang, Tiesong Zhao, \emph{Senior Member, IEEE}, and Xiaoming Tao, \emph{Senior Member, IEEE}

\thanks{This work is supported by the National Natural Science Foundation of China (No. 62571131). {\it(Shuijie Li and Kemi Chen contributed equally to this work.)(Corresponding author: Tiesong Zhao.)}

S. Li, K. Chen, R. Wang and T. Zhao are with the Fujian Key Laboratory for Intelligent Processing and Wireless Transmission of Media Information, College of Physics and Information Engineering, Fuzhou University, Fuzhou 350108, China, and  Fujian Science \& Technology Innovation Laboratory for Optoelectronic Information of China, Fuzhou 350108, China. (e-mails: 231120041@fzu.edu.cn, arice\_chen@163.com, wangrunjie2023@163.com, t. zhao@fzu.edu.cn).

X. Tao is with the Department of Electronic Engineering, Tsinghua University, Beijing 100084, China (e-mail: taoxm@tsinghua.edu.cn)}}


\maketitle

\begin{abstract}
The increasing demand for real-time tactile communication in multimedia systems has exposed the limitations of existing Joint Source-Channel Coding (JSCC) techniques. While current JSCC models facilitate end-to-end optimization, they typically operate at fixed coding rates and require separate model instances for different rate settings. This results in significant storage overhead and limited adaptability to dynamic bandwidth conditions. To address these challenges, we propose the Flexible Deep Joint Source-Channel Coding (FD-JSCC) framework for vibrotactile signals, which supports flexible-rate transmission without the need for model switching. The FD-JSCC integrates a flexible-rate encoder-decoder enhanced with Hierarchical Gain Adaptation Module (HGAM) and Rate-Switchable Residual Module (RSRM), enabling bitrate-aware compression by selectively preserving salient vibrotactile features. Additionally, we introduce a Channel Feature Processing Module (CFPM), which leverages real-time SNR information to enhance robustness against channel noise and signal degradation. Trained on the IEEE 1918.1.1 vibrotactile dataset, FD-JSCC achieves reconstruction performance comparable to fixed-rate baselines (e.g., DeepSC-S), while reducing storage requirements by 61.1\% when supporting four rates. These results underscore its potential for scalable, low-latency tactile communication in next-generation networks.
\end{abstract}


\section{Introduction}

\begin{figure}[!t]
    \centering
    \includegraphics[width=1\linewidth]{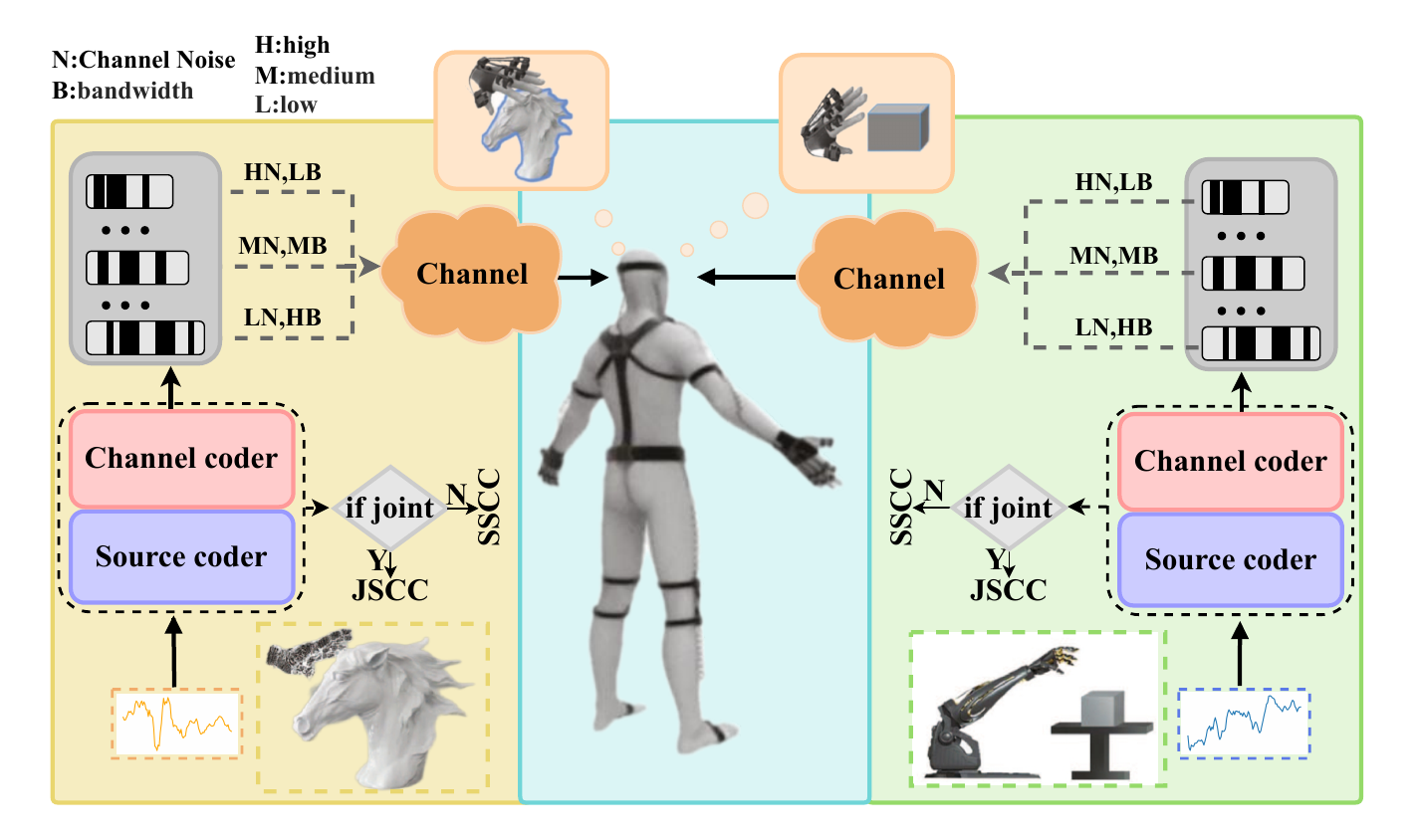}
    \caption{ The immersion and telepresence of tactile information.}
    \label{fig1}
\end{figure}

\IEEEPARstart{T}{actile} perception is essential for humans to interact effectively with objects and their environment~\cite{silva2019subjective}. To enhance the sense of immersion in Virtual Reality (VR) and teleoperation, systems typically employ vibrotactile feedback to improve the user experience~\cite{zhu2020multimedia}, thereby increasing the realism of VR scenarios and teleoperation~\cite{froehlich2024can}. As illustrated in Fig.~\ref{fig1}, when users engage in a virtual reality scenario or perform remote control, vibrotactile information is collected at the slave end and transmitted back to the master end. This requires a communication system with low latency and high transmission efficiency to ensure that users can achieve an immersive experience.

Communication systems typically consist of two key components: source coding, which reduces redundancy in the source data, and channel coding, which minimizes or corrects transmission errors to enhance data reliability. Traditional communication system designs treat source and channel coding independently, without considering the impact of other modules. This separated architecture often results in long transmission delays, rendering it unsuitable for high-real-time, low-latency tactile communication scenarios. Furthermore, from an end-to-end optimization perspective, this separated architecture is a suboptimal choice. In contrast, Joint Source-Channel Coding (JSCC) achieves global optimality by jointly optimizing both source and channel coding, thereby reducing latency and enhancing noise robustness—especially in low Signal-to-Noise Ratio (SNR) environments. This approach effectively addresses the high latency and ``cliff effect" (drastic performance degradation under channel degradation) associated with Separate Source-Channel Coding (SSCC), making it ideal for real-time applications such as tactile communication.


The SSCC paradigm is based on Shannon’s three-stage coding theory. Within this framework, source encoders (e.g., Huffman or arithmetic coding) aim to eliminate statistical redundancy, while channel encoders (e.g., LDPC or Polar codes) address transmission noise. However, in tactile signal transmission, SSCC faces three significant limitations. First, its modular design hinders global optimization; the distinct objectives of redundancy reduction and error correction often conflict, particularly due to the non-stationary characteristics of tactile signals~\cite{tung2022deepjscc}. Second, SSCC systems are highly vulnerable to the cliff effect, where performance deteriorates rapidly when SNR falls below a certain threshold—an issue exacerbated by the time-varying nature of 5G millimeter-wave channels~\cite{huang2025d,bourtsoulatze2019deep}. Third, traditional methods operate at the symbol level and fail to preserve high-level semantic features inherent in tactile signals, such as the spatial distribution of saliency in force-feedback patterns, which disconnects transmission efficiency from perceptual quality~\cite{weaver1953recent}.


Although joint source–channel coding (JSCC) has been extensively investigated for images~\cite{bourtsoulatze2019deep,zhang2023adaptive,erdemir2023generative,kurka2021bandwidth,xie2024deep,jarrahi2024dcs,huang2024joint,sadhu2023high,yang2024swinjscc,li2024content,han2025scsc,qi2024g,yi2024joint,bao2021adjscc,jarrahi2024joint,huang2023mi,chen2023deep}, video~\cite{argyriou2008error}, speech~\cite{bokaei2023deep,bokaei2025low,weng2021semantic,bokaei2024deep}, and text~\cite{xie2021deep}, its application to tactile signals~\cite{10829930} remains relatively nascent. Existing tactile-focused studies, including recent work on multi-channel vibrotactile sensors, have primarily targeted compression and representation; they seldom consider real-time transmission over time-varying wireless channels or the need to adapt to changing SNR. In particular, most current JSCC models do not support variable coding rates, which limits their flexibility across heterogeneous deployment scenarios with differing bandwidth and latency constraints. To address these gaps, we propose a flexible JSCC framework that dynamically adjusts its coding rate according to channel conditions, enabling robust, high-quality real-time tactile feedback in unstable wireless environments. Our proposed method is evaluated using the IEEE 1918.1.1 standard vibrotactile dataset~\cite{kirsch2018low}. The contributions of this paper can be summarized as follows:

\begin{itemize}

\item[$\bullet$]

We propose a novel JSCC framework for vibrotactile signals that supports flexible output rates. This design allows for adjustments to the coding rate, enabling users to balance model performance with application-specific requirements.

\item[$\bullet$]

We introduce a Channel Feature Processing Module (CFPM). Unlike conventional separate source-channel coding schemes that suffer from the cliff effect, as well as existing deep JSCC methods with limited generalization capabilities, the proposed CFPM extracts channel-aware features from both discrete and continuous perspectives. This enables the model to dynamically adapt to varying channel conditions and improves its robustness across a wide range of SNR levels.

\item[$\bullet$]

Our method achieves state-of-the-art performance on a standard dataset, attaining a ST-SIM of 97.8\% at a 10 dB SNR and a compression ratio of 2, surpassing existing algorithms in both compression efficiency and perceptual quality. Furthermore, it satisfies the low-latency requirements that are essential for vibrotactile applications.

\end{itemize}

The remainder of this paper is organized as follows: Section~\ref{Related Work} reviews existing literature on JSCC and tactile data compression. Section~\ref{Proposed Method} offers a comprehensive description of the proposed framework. Section~\ref{Experiments and Discussions} presents the experimental results along with their analysis. Finally, Section~\ref{Conclusion} summarizes the key findings of the paper.

\section{Related Work}
\label{Related Work}

\subsection{Joint Source-channel Coding (JSCC)}
As a pivotal technology for addressing the limitations of traditional communication, JSCC has attracted significant research interest from numerous scholars in recent years, resulting in substantial advancements in both theoretical exploration and practical applications. Liu \textit{et al.}~\cite{liu2020joint} proposed (Multi-Objective Differential Evolution) MODE-based optimization for double protograph low-density parity-check JSCC systems, co-designing source and channel codes along with their interconnections to reduce decoding thresholds and complexity for efficient implementation. This algorithm comprehensively considers various factors, including source distribution, coding rate, and code length, while jointly optimizing system components such as source codes, channel codes, and connection edges. The Nonlinear Transformation Source-Channel Coding (NTSCC) introduced by Dai \textit{et al.}~\cite{dai2022nonlinear} innovatively integrates nonlinear transformation with deep JSCC. By establishing a learnable entropy model, NTSCC can accurately approximate the source distribution, providing effective guidance for subsequent coding decisions. Wang \textit{et al.}~\cite{wang2022wireless} designed a deep video semantic transmission framework that employs nonlinear transformation and a conditional coding architecture to adaptively extract the semantic features of video frames. Huang \textit{et al.}~\cite{huang2025d} proposed a digital-depth JSCC framework that seamlessly combines deep source coding with digital channel coding. Through approximate analysis of end-to-end distortion, this framework can optimize both source and channel rates. Kurka \textit{et al.}~\cite{kurka2020deep} introduced a deep JSCC scheme based on autoencoders, which fully leverages the feedback mechanism to enhance the performance of image transmission. Additionally, some studies focus on integrating JSCC with emerging technologies, such as 5G and 6G communication systems, to meet the demands for high-speed and low-latency communication.
 
 \subsection{Tactile Compression}
 
 JSCC schemes dedicated to tactile data transmission remain limited. Current research in tactile data primarily focuses on source compression coding. Zhao \textit{et al.}~\cite{zhao2022high} proposed RNVC, which is based on gated recurrent units. This codec fully exploits the statistical characteristics of vibrotactile data. While ensuring perceptual quality, it significantly enhances compression efficiency and reduces encoding delay. In the realm of perception-driven compression strategies, Hassen \textit{et al.}’s~\cite{hassen2020pvc} PVC-SLP designs frequency-domain quantization thresholds, preserves key perceptual features, and outperforms traditional wavelet transforms in low-bitrate scenarios, providing valuable insights for resource-constrained conditions. Regarding algorithm improvement and performance enhancement, Noll \textit{et al.}’s~\cite{noll2021vc} VC-PWQ optimizes the perceptual model, quantization method, and arithmetic coding to enhance compression efficiency, thereby meeting the demands of low-data-rate transmission in multi-point interaction scenarios. Additionally, inspired by embodied artificial intelligence, Lu \textit{et al.}~\cite{lu2025cross} designed a cross-modal tactile compression scheme. This scheme extracts visual semantic features through the multi-dimensional tactile feature fusion network (MTFFN) to guide the parameter optimization of the tactile codec, thereby improving perceptual quality. Xu \textit{et al.}’s NNVC~\cite{xu2024efficient}, an Nbeats-based encoder-decoder, integrates logarithmic quantization and entropy coding, surpassing existing technologies in both compression ratio and signal quality metrics.

Overall, research in vibrotactile coding is advancing across multiple dimensions—signal modeling, perceptual optimization, dataset development, and algorithm design—establishing a foundation for low-latency, high-fidelity transmission in multi-channel interactive systems. These advancements support a wide range of applications in virtual reality, teleoperation, and smart healthcare. Looking ahead, JSCC techniques for tactile signals show promise for emerging scenarios such as telesurgery and virtual reality training, enhancing the quality of human–machine interaction. As communication technologies evolve, the demand for immersive sensory experiences grows~\cite{simiscuka2023omniscent}. As a key modality alongside vision and hearing, touch facilitates multi-level stimulation and more realistic interactions~\cite{xu2019error,deng2023long}, garnering increasing attention across various fields. However, a review of current tactile-data JSCC studies indicates a strong emphasis on compression, with limited focus on real-time transmission challenges in dynamic, noisy channels. Most existing one-dimensional JSCC methods lack flexible-rate capabilities, which are crucial for adapting to fluctuating channel conditions. Such adaptability enables high-rate, high-fidelity transmission under optimal conditions while ensuring robustness and continuity in the face of bandwidth constraints.

\begin{figure}[t]
    \centering
    \includegraphics[width=2.5in]{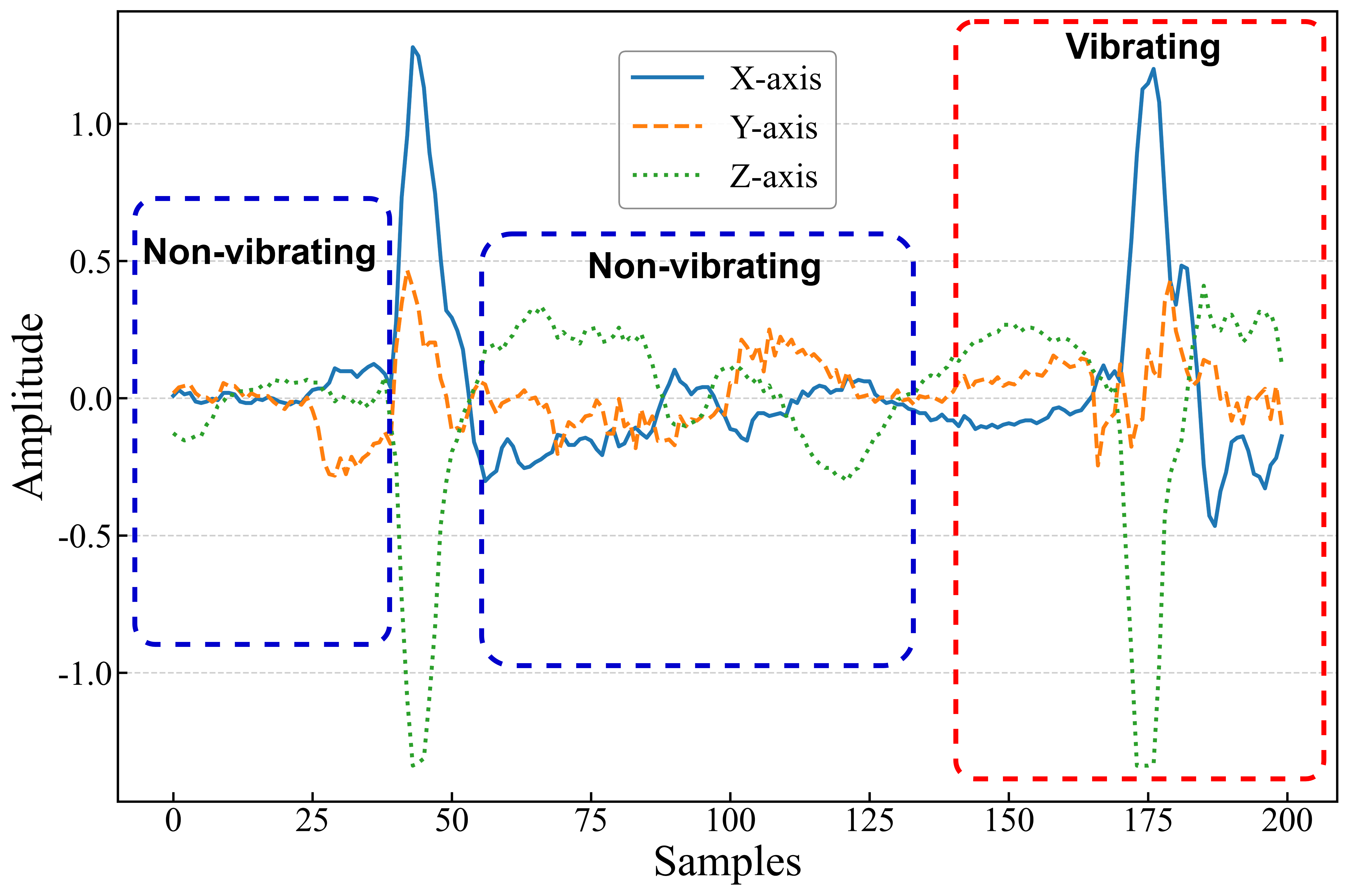}
    \caption{A Typical vibrotactile signal.}
    \label{figyl}
\end{figure}

\section{Proposed Method}
\label{Proposed Method}
\subsection{Task description}


To address the limitations of fixed-rate coding, we propose a framework for FD-JSCC inspired by classical speech coding techniques, with redundancy removal as a core strategy. This approach targets temporal, spectral, and perceptual redundancies within the signal. Similarly, vibrotactile signals exhibit both temporal and perceptual redundancies (see Fig.~\ref{figyl}); periods of near-zero amplitude represent a “non-vibrating” state, analogous to silence in speech. Furthermore, considering the limited sensitivity of human tactile perception—particularly to low-amplitude and high-frequency components—many segments of the signal can be eliminated with minimal perceptual impact.

These near-zero amplitude intervals indicate temporal redundancy in tactile signals. Given that human tactile perception is relatively insensitive to subtle vibrations, perceptual spectral redundancy can also be effectively minimized. By selectively eliminating both types of redundancy, we establish the following decomposition:
\begin{align}
\label{yl}
\mathrm{F_V=F_I+F_R},
\end{align}
where $\mathrm{F_V}$ denotes the complete set of vibrotactile features extracted from the original signal, this set can be divided into two subsets: essential vibrotactile features $\mathrm{F_I}$, and redundant or non-essential features $\mathrm{F_R}$. The features in $\mathrm{F_I}$ carry higher semantic and perceptual significance compared to those in $\mathrm{F_R}$. To support flexible rate transmission, we adaptively retain a variable number of features from $\mathrm{F_V}$ according to the target bit rate, with $\mathrm{F_I}$ representing the retained subset.
\begin{align}
\begin{split}
\label{yl1}
&\mathrm{F_V=\left \{F_{V1},F_{V2},F_{V3},...,F_{Vn}  \right \}},\\
&\mathrm{F_I=\left \{F_{V1},F_{V2},F_{V3},...,F_{Vm}  \right \}},\\
&\mathrm{F_R=\left \{F_{Vm+1},F_{Vm+2},F_{Vm+3},...,F_{n}  \right \}}.
\end{split}
\end{align}
\begin{figure}[t]
    \centering
    \includegraphics[width=2.5in]{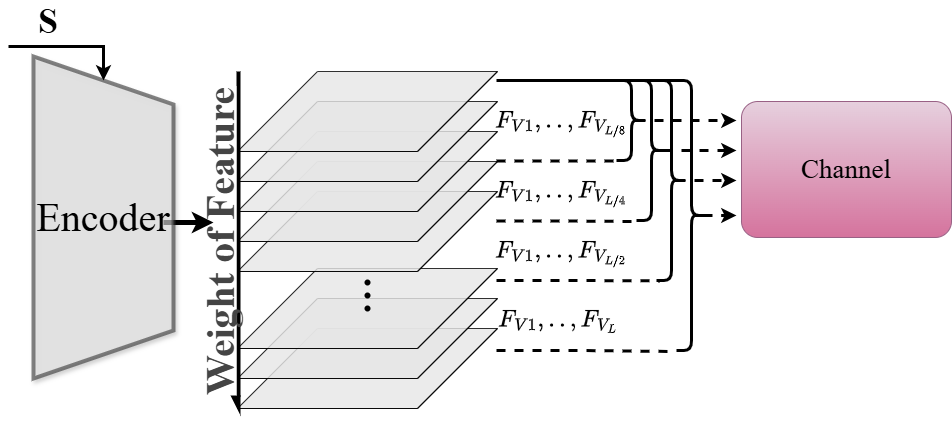}
    \caption{Select the number of features for encoding output.}
    \label{fffig}
\end{figure}
We assume that the features $\mathrm{F_V}$ are ranked by importance, from high to low. Based on the target bit rate, we can adjust the number of retained features ($\mathrm{F_I}$). We do this by selecting the top-$m$ elements. This approach enables us to efficiently balance transmission bandwidth and perceptual fidelity. It is noteworthy that in the subsequent implemented architecture (Section III.B, Fig.~\ref{fig2}), this conceptual ``selection" of the feature set $F_I$ is not a separate pre-processing step. Instead, it is intrinsically realized by the learnable modules (HGAM, RSRM, and RSCM) under the control of the parameter S, which collectively work to prioritize and compress the most salient features into the transmitted representation.

We specifically assume that the features $\mathrm{F_V}$ are ranked in descending order of importance. Based on the target bit rate, we determine the number of features to retain by selecting the top-$m$ elements to form $\mathrm{F_I}$. This approach facilitates an efficient trade-off between transmission bandwidth and perceptual fidelity, akin to traditional speech coding, which preserves critical spectral components while discarding redundant information. As illustrated in Fig.~\ref{fffig}, the output code rate is controlled by adjusting the number of features fed into the encoder. A larger number of input features results in a higher output code rate, whereas fewer input features yield a lower code rate.

\subsection{Overall JSCC Framework}

\begin{figure*}[!t]
    \centering
    \includegraphics[width=6.5in]{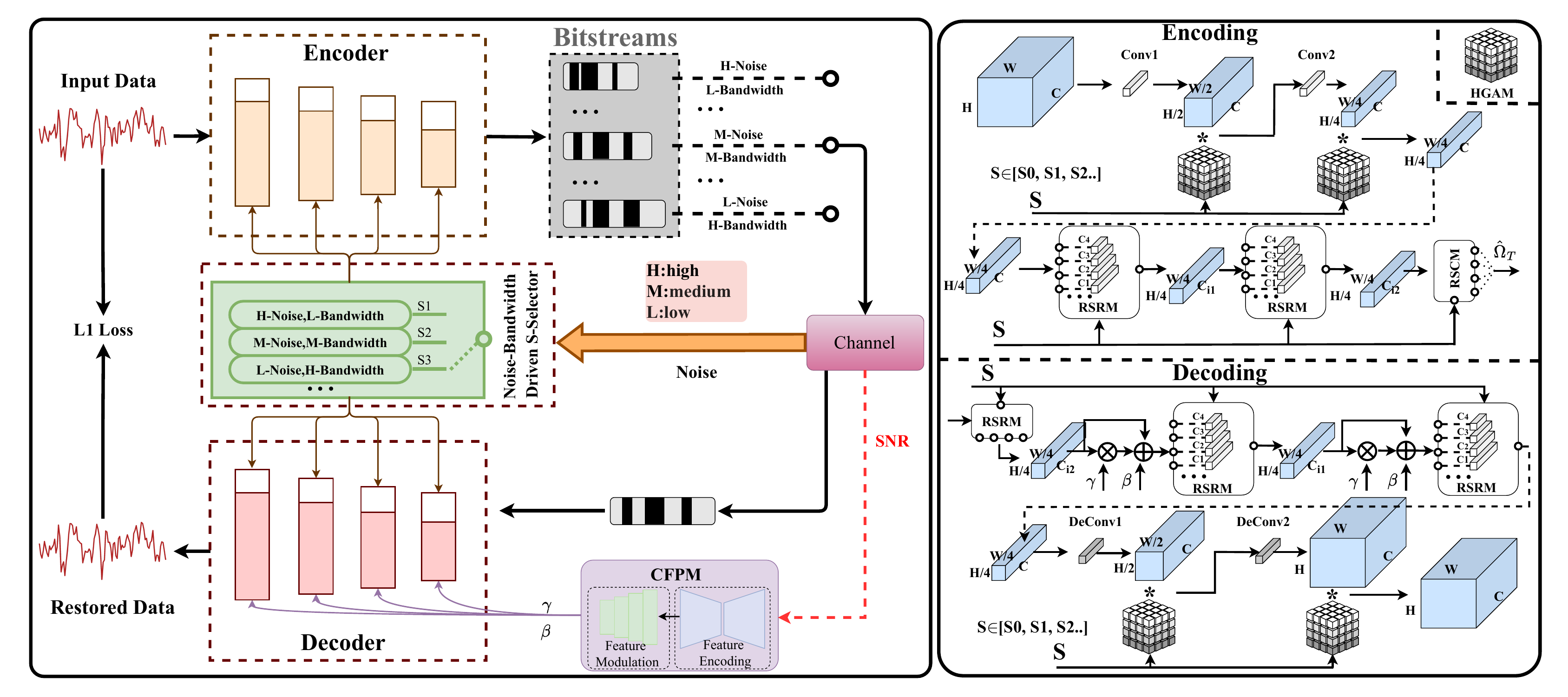}
    \caption{The proposed Flexible Deep Joint Source-Channel Coding (FD-JSCC) framework.}
    \label{fig2}
\end{figure*}

As illustrated in Fig.~\ref{fig2}, we propose an innovative JSCC-based coding framework for vibrotactile data. The core design emphasizes a flexible-rate coding mechanism, developed based on the principles previously mentioned for achieving flexible-rate control. This mechanism utilizes an end-to-end architecture that jointly optimizes both the encoder and decoder, enabling dynamic adaptability to varying channel conditions.

A key feature of this framework is the parameterized design of switchable-rate modules, which enables users to adjust the compression ratio in real time by selecting a control quantization parameter \(S \in [0, 1, 2, 3]\). This design eliminates the need to reload model weights, thereby facilitating seamless adaptation to varying bandwidth scenarios. To enhance robustness against channel noise, the model incorporates a CFPM that extracts channel noise characteristics in real time and modulates the encoding process accordingly. This design ensures that perceptually critical vibrotactile information is preserved even under low bit-rate constraints.

\begin{figure}[t]
    \centering
    \includegraphics[width=2.5in]{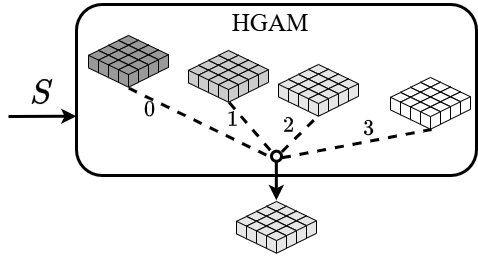}
    \caption{The structural diagram of the Hierarchical Gain Adaptation Module (HMGA). The HGAM generates a corresponding learnable Hadamard matrix by utilizing various parameters $S$, which facilitates the entire network's ability to adapt more rapidly across different compression ratios.}
    \label{fig3}
\end{figure}

\textbf{Problem formulation.} Given an input signal sequence $\mathbf{x} = \{x_1, x_2, ..., x_N\}$, different coding strategies with varying bit rates are selected via the parameter $S$, and the reconstructed signal is denoted as $\hat{\mathbf{x}}^S$. The L1 distortion is defined as:
\begin{align}
L_1(\mathbf{x}, \hat{\mathbf{x}}^S) = \frac{1}{N} \sum_{i=1}^{N} |x_i - \hat{x}_i^S|.
\end{align}
\textbf{Optimization Objective.} Minimize the L1 distortion for different values of $S$.
\begin{align}
\arg\min_{S \in \{0,1,2,3\}} L_1(\mathbf{x}, \hat{\mathbf{x}}^S).
\end{align}
Now, we present a detailed description of our algorithm. We adjust the model's compression ratio through a parameterized design by controlling the parameter $S\in \left [ 0,1,2,3 \right ]$(). Meanwhile, we allow the vibrotactile signals to pass through the Vibrotactile Features Extraction Module. Next, we perform a Hadamard product operation on the output with the HGAM, which is controlled by $S$, which retains the critical vibrotactile features (as shown in Fig.~\ref{fig3}). This process enhances the model's fitting speed. Subsequently, we pass the vibrotactile features through two Rate-Switchable Residual Modules (RSRMs) with independent parameters, enabling us to further extract essential vibrotactile features $\Omega_2$. Finally, we input these features into the RSCM for transmission, where BN denotes Batch Normalization~\cite{ioffe2015batch}.
\begin{align}
\begin{split}
\label{HGAM}
&\Omega =\mathrm{BN}(\mathrm{Conv_1}(X))\odot \mathrm{HGAM_1}(S),\\
&\Omega_1 =\mathrm{BN}(\mathrm{Conv_2}(\Omega))\odot \mathrm{HGAM_2}(S).
\end{split}
\end{align}
\begin{table}[t]
\caption{The structure of the Rate-Selective Channel Module, where S is the input parameter.\label{table_DCR}}
\centering
\begin{tabular}{cc}
\hline
S & Rate-Selective Channel Module(RSCM)\\
  & Conv(in,out,k) \\
  \hline
 & in:Input channels\\
S $\in\{0,1,2,3\}$ & out:Output channels\\
 & k:Convolution kernel size\\
  \hline
0 & Conv(128,2,5),BNLayer\\
1 & Conv(128,4,5),BNLayer\\
2 & Conv(128,8,5),BNLayer\\
3 & Conv(128,16,5),BNLayer\\
\hline
\end{tabular}
\end{table}
The RSCM is a structurally simple yet effective component. It selects convolutional layers with varying numbers of output channels based on the parameter $S$. The detailed architecture of RSCM is presented in Table~\ref{table_DCR}. This module enables further compression of vibrotactile features and transforms them into symbols $\hat{\Omega}_T$ that are suitable for transmission over the communication channel. This transmitted representation $\hat{\Omega}_T$ effectively constitutes the final, rate-adapted realization of the essential feature set $F_I$.
\begin{align}
\label{DCR-CS}
\hat{\Omega}_\mathrm{T} = \mathrm{BN}(\mathrm{RSCM_1}(\Omega_2, S)).
\end{align}
Then, we transmit $\hat{\Omega}_T$ through the communication channel. In this study, we consider a Gaussian channel, where Gaussian white noise $N$ is added to the signal. The decoder receives the transmitted symbols and reconstructs the original information. The architecture of the decoder is approximately symmetric to that of the encoder; however, a key distinction lies in the incorporation of prior channel knowledge, specifically the channel SNR, during the decoding process. This is achieved through the CFPM, which extracts channel-aware features based on prior SNR information. These features are subsequently integrated into multiple layers of the decoder to modulate the internal feature representations. By incorporating the CFPM, the decoder adaptively adjusts its representations layer by layer according to channel conditions, improving reconstruction quality under varying SNR levels.

\subsection{Analysis of vibrotactile characteristics}
As illustrated in Fig.~\ref{figyl}, the vibrotactile signals exhibit a wide amplitude range, spanning approximately from $-1.34$ to $+1.57$, and include both positive and negative values. This reflects the bidirectional nature of vibration signals, consistent with their inherent directional reversals. Certain regions display densely distributed and rapidly varying values, indicating the presence of high-frequency components. The data points demonstrate a periodic fluctuation pattern, characterized by alternating peaks and troughs. In some intervals, values change gradually, exhibiting smooth transitions from negative to positive, while in others, they display sharp, pulse-like variations. The X, Y, and Z axes share a similar global trend, although differences in amplitude and phase are evident across the axes. Overall, the signal encompasses variations at multiple temporal scales, combining fast oscillatory components with slower envelope modulations.
\begin{figure*}[t]
    \centering
    \includegraphics[width=1\linewidth]{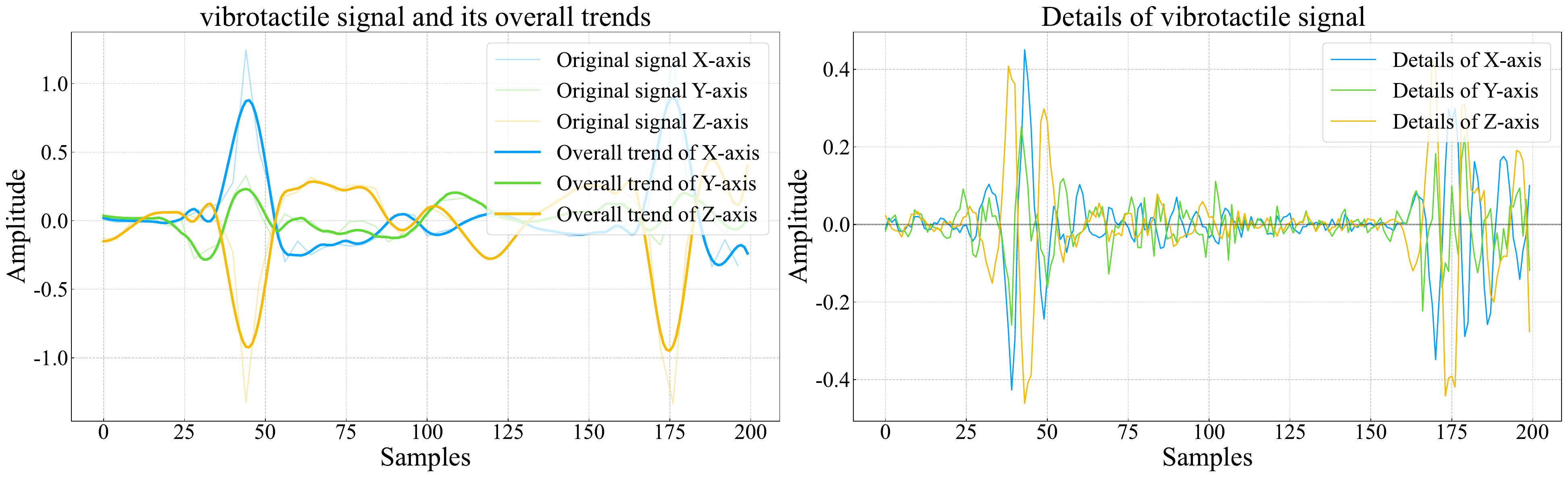}
    \caption{Vibrotactile Signal Analysis.}
    \label{figyvla}
\end{figure*}
Meanwhile, we apply Savitzky-Golay filtering to the original signal to visualize its slow envelope and fast oscillating components, as shown in Fig. 6, for illustrative purposes only. This preprocessing step aids in visualizing the multi-scale temporal structure but is not involved in the perceptual feature selection process of our model.

\subsection{Network Design}
In the previous subsection, we established that the range of vibrotactile signal values is extensive, and these signals are susceptible to transmission noise, with variations in SNR significantly impacting signal quality. To address this issue, we propose the Channel Feature Processing Module (CFPM), which simultaneously performs channel feature extraction and modulation to enhance robustness against channel degradation.

Specifically, the CFPM processes the input channel SNR through two complementary branches: a discretization-based embedding pathway and a continuous-value transformation pathway. In the first branch, the SNR is discretized into bins and passed through an embedding layer to learn non-linear correlations across different SNR intervals, resulting in the dense vector $\hat{N}_1$. In parallel, the raw SNR is processed by a multilayer perceptron, followed by an SNR-aware attention mechanism, to generate the refined feature vector $\hat{N}_2$. These two vectors are then combined to obtain the final channel feature representation:\begin{align}
\label{CA-CRB1}
\begin{split}
& \hat{N}_1 =\mathrm{E}[\mathrm{\phi} (\mathrm{SNR})],\\
& \hat{N}_2 = \mathrm{Att}[\mathrm{MLP}(\mathrm{SNR})],\\
& \hat{N} = \hat{N}_1 +\hat{N}_2.
\end{split}
\end{align}
Among them, $\mathrm{E[\cdot]\in K\times d}$ is the embedding matrix, $\mathrm{\phi \in [1,2,3,...K]}$ represents the discretization of SNR into K bins, $\mathrm{MLP(\cdot)}$ denotes the non-linear transformation, and $\mathrm{Att[\cdot]}$ is the attention mechanism.

Next, the CFPM modulates the encoded signal features based on $\hat{N}$. Specifically, $\hat{N}$ is passed through two independent fully connected layers followed by nonlinear activations to generate a channel-wise scaling factor $\gamma$ and bias term $\beta$. Finally, modulation is applied to the encoded feature $\hat{Z}$ as:
\begin{align}
\label{CA-CRB2}
\begin{split}
&\gamma = \mathrm{\delta}(\mathrm{W}_{\mathrm{\gamma}}\hat{N}+b_{\mathrm{\gamma}}),\\
&\beta = \mathrm{tanh}(\mathrm{W}_{\beta}\hat{N}+b_{\beta}),\\
&\hat{Z}_1 = \hat{Z}\times(\gamma+1)+\beta.
\end{split}
\end{align}
where $\delta(\cdot)$ is the sigmoid activation, and $\tanh(\cdot)$ denotes the hyperbolic tangent function.

This modulation allows the model to dynamically adapt to changing channel conditions by selectively emphasizing or suppressing features. This approach ensures the preservation of perceptually critical vibrotactile information, even in scenarios with low SNR or compression constraints. Given that vibrotactile signals exhibit variations across multiple temporal scales—encompassing both rapid oscillatory components and slowly varying envelopes, as illustrated in Fig.~\ref{figyvla}—we propose the RSRM to effectively capture these multi-scale characteristics. The RSRM employs a hierarchical convolutional architecture for multi-scale feature extraction, incorporates residual connections to preserve the global structure of the input signal, and integrates a Squeeze-and-Excitation (SE) attention mechanism to adaptively reweight frequency components. This design enables the model to emphasize perceptually significant vibratory features while maintaining the overall integrity of the signal.
\begin{align}
\label{CA-CRB0}
\begin{split}
& \varphi_{\mathrm{out}} = \mathrm{RSRM}(S, \varphi_{\mathrm{in}),
}\\
& \hat{\varphi} = \mathrm{\sigma} (\mathrm{BN}(\mathrm{RSCM_1}(S,\varphi_{\mathrm{in}}))),\\
& \varphi_{\mathrm{out}} = \varphi_{\mathrm{in}}+\mathrm{G}(\mathrm{BN}(\mathrm{RSCM_2}(S,\hat{\varphi}))).
\end{split}
\end{align}
Here, $\mathrm{F(\cdot)}$ represents the RSCM, $\mathrm{\sigma(\cdot)}$ represents the ReLU layer, and $\mathrm{G(\cdot)}$ represents the SE module.
\section{Experiments and Discussions}
\label{Experiments and Discussions}

\subsection{Experiment Setup}


To evaluate our method, we implemented it in Python and conducted experiments on a PC equipped with an NVIDIA GeForce RTX 3090 to compare its performance against other popular approaches. We utilized the standard dataset and testing conditions established by the IEEE P1918.1.1 working group to ensure fair comparisons. Divide the dataset into a training set and a test set for training and testing. This dataset comprises 280 vibrotactile signals recorded under various materials and exploration speeds, with each signal consisting of 3,360 to 8,400 samples represented by 16-bit pulse codes. The sampling rate is 2,800 Hz, which corresponds to an original bit rate of 44.8 kbps. As recommended, all signals were converted to a one-dimensional format. Direct training on this dataset led to underfitting; so we split each one-dimensional signal into multiple segments of 144 samples for training. The parameter $Q$ controls the scaling factor, which we set to 1000 as specified below.


In this study, we conducted a comprehensive comparative analysis of coding algorithms for vibrotactile signals, specifically RNVC~\cite{zhao2022high}+16QAM and VC-PWQ~\cite{noll2021vc}+16QAM. Furthermore, we evaluated the JSCC frameworks designed for speech transmission scenarios, including the DeepSC-S algorithm~\cite{weng2021semantic}, DNN algorithm~\cite{bokaei2023deep}, and DNN2 algorithm~\cite{bokaei2025low}. All codecs were tested using a complete standard dataset. To quantitatively assess the performance of the codecs, we introduced the following evaluation criteria: the compression ratio (CR) to measure compression performance, the peak signal-to-noise ratio (PSNR) to evaluate signal reconstruction quality, the perception-based reconstruction quality metric ST-SIM~\cite{hassen2019subjective}, and the Perceptual Mean Square Error (PMSE)~\cite{chaudhari2011towards}. Additionally, we assessed PSNR, PMSE, and ST-SIM under varying channel SNRs to evaluate the models' resilience against channel noise. The compression ratio is defined as the ratio of the original bitrate to the compressed bitrate.

\subsection{Model Parameters}



\begin{figure*}[!t]
    \centering
    \includegraphics[width=1\linewidth]{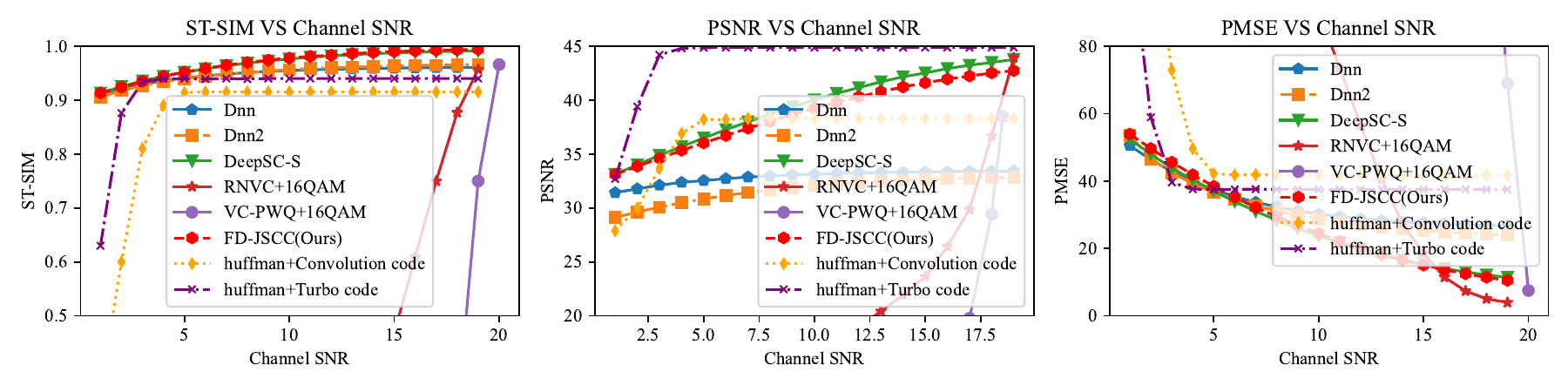}
    \caption{The comparative performance evaluation of the proposed FD-JSCC framework was conducted against DNN, DNN2, DeepSC-S, RNVC+16QAM, VC-PWQ+16QAM, huffman+Convolution, and huffman+Turbo, in terms of ST-SIM, PSNR, and PMSE, and was carried out under varying channel SNR conditions with a fixed CR of 2.}
    \label{fig7}
\end{figure*}

\begin{table}[!t]
\caption{The number of parameters for different algorithms. \label{parameters}}
\centering
\begin{tabular}{c|c|c|c|c|c}
\hline
\multicolumn{6}{c}{Number of parameters (M)} \\ \hline
Algorithm & CR = 2 & CR = 4 & CR = 8 & CR = 16 & Total \\ \hline
DNN & 2.21M & 1.10M & 0.63M & 0.37M & 4.31M \\

DNN2 & 0.21M & 0.21M & 0.21M & 0.21M & 0.84M \\

DeepSC-S & 5.06M & 5.01M & 4.98M & 4.97M & 20.02M \\ \hline
Ours & \multicolumn{4}{c|}{7.80M} & 7.80M \\ \hline
\end{tabular}
\end{table}
\begin{table}[!t]
\caption{Comparison of model delays.\label{del}}
\centering
\begin{tabular}{c|c|c|c|c}
\hline
\multicolumn{5}{c}{Encoding and decoding latency of models (ms)} \\ \hline
Model & DNN & DNN2 & DeepSC-S & Ours \\ \hline
Delay & 0.7 & 5.2 & 3.8 & 2.5 \\
\hline
\end{tabular}
\end{table}
As shown in Table~\ref{parameters}, for the DNN, DNN2, and DeepSC-S algorithms, using different compression ratios (e.g., $\mathrm{CR} \in {2, 4, 8, 16}$, corresponding to coding rates of $\mathrm{3276.8,kbps}$, $\mathrm{1638.4,kbps}$, $\mathrm{919.2,kbps}$, and $\mathrm{409.6,kbps}$, respectively) requires switching to separately trained model weights. In contrast, our model supports multi-rate output through a single model. Users only need to adjust the parameter $S$ to select the desired compression ratio, offering greater flexibility and ease of use. Meanwhile, as shown in Fig.~\ref{fig7}, our algorithm significantly outperforms both the DNN and DNN2 algorithms in terms of performance, and its performance is comparable to that of the DeepSC-S algorithm. However, in comparison to the DeepSC-S model, our model reduces the total number of parameters by approximately 61.03$\%$.

\begin{figure*}[!t]
    \centering
    \includegraphics[width=1\linewidth]{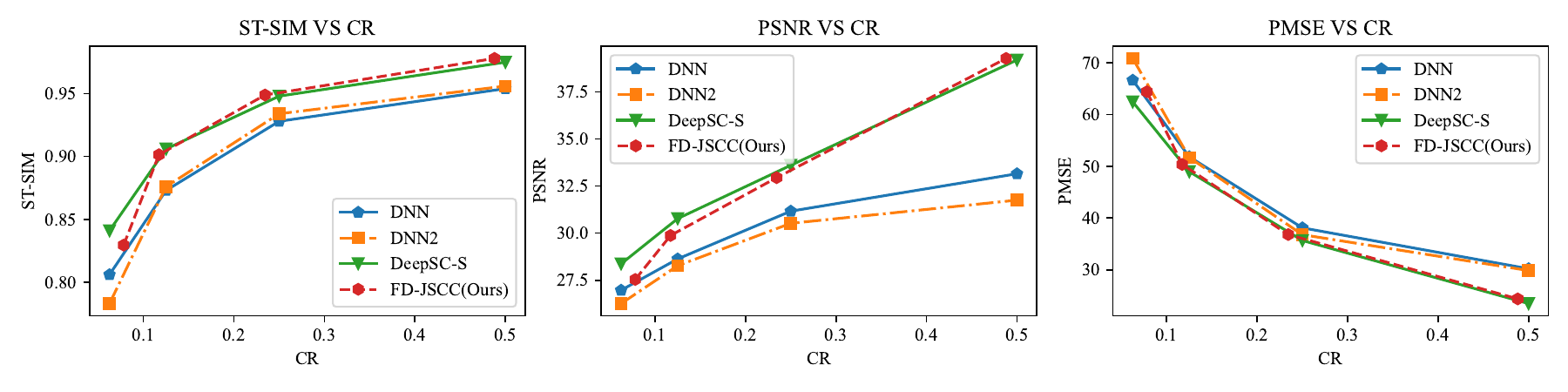}
    \caption{The comparative performance evaluation of the proposed FD-JSCC framework is conducted against DNN, DNN2, and DeepSC-S in terms of ST-SIM, PSNR and PMSE, under a fixed channel SNR of 10 dB across various CRs.}
    \label{fig8}
\end{figure*} 
\begin{figure}[!t]
    \centering
    \includegraphics[width=1\linewidth]{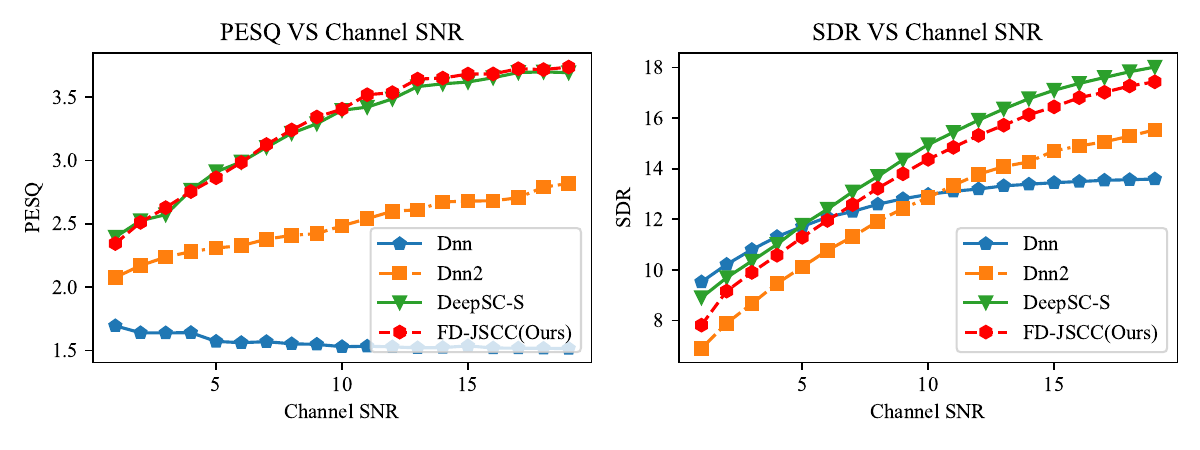}
     \caption{The comparative performance evaluation of the proposed FD-JSCC framework was conducted against DNN, DNN2 and DeepSC-S in terms of PESQ and SDR, conducted under varying channel SNR conditions at a fixed CR of 2.}
    \label{fig9}
\end{figure} 
\begin{figure}[!t]
    \centering
    \includegraphics[width=1\linewidth]{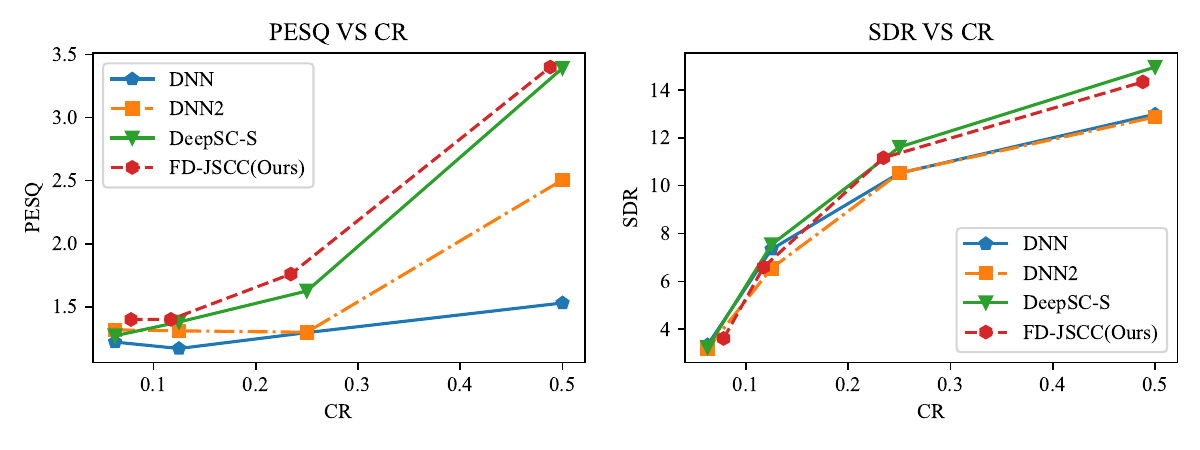}
    \caption{The comparative performance evaluation of the proposed FD-JSCC framework was conducted against DNN, DNN2, and DeepSC-S in terms of PESQ and SDR, under a fixed channel SNR of 10 dB across various CRs.}
    \label{fig10}
\end{figure}
\subsection{Codec Delay}

As shown in Table~\ref{del}, the algorithmic latency of our proposed model is 2.5 ms, which is well below the perceptual threshold, thereby meeting the requirements for real-time tactile applications. Furthermore, in terms of processing speed, our model achieves faster encoding and decoding compared to DNN2 and DeepSC-S, while demonstrating slightly slower performance than DNN.
\subsection{Model Performance}
From Fig.~\ref{fig7}, due to the well-known ``cliff effect", the RNVC + 16QAM scheme exhibits degraded performance under low channel SNR conditions. In contrast, the performance curves of the other models remain smooth within the SNR range of [1, 19], where the CR for all models is uniformly set to 2. Our proposed model significantly outperforms both DNN and DNN2 across all three evaluation metrics—ST-SIM, PSNR, and PMSE—as the channel SNR varies. Compared to DeepSC-S, our model achieves slightly better performance in terms of ST-SIM and PMSE, while exhibiting marginally lower performance in PSNR. These results demonstrate that our model consistently achieves competitive performance across different levels of channel noise. Notably, when the channel SNR exceeds 15 dB, the ST-SIM score of our model surpasses 99$\%$.
\begin{figure*}[!t]
    \centering
    \includegraphics[width=1\linewidth]{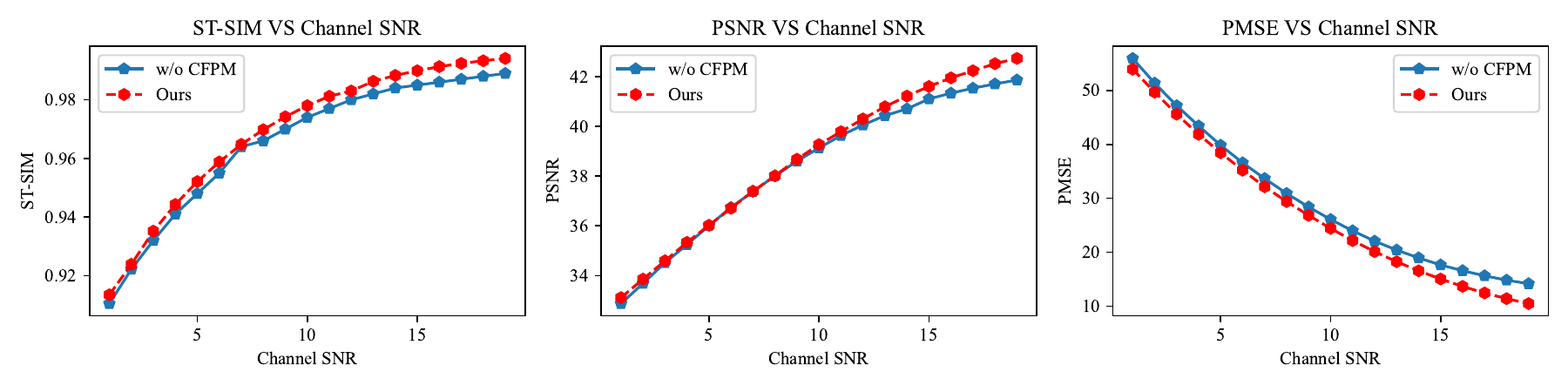}
     \caption{The performance of the model with and without CFPM under different channel SNR.}
    \label{fig11}
\end{figure*} 
\begin{figure*}[!t]
    \centering
    \includegraphics[width=1\linewidth]{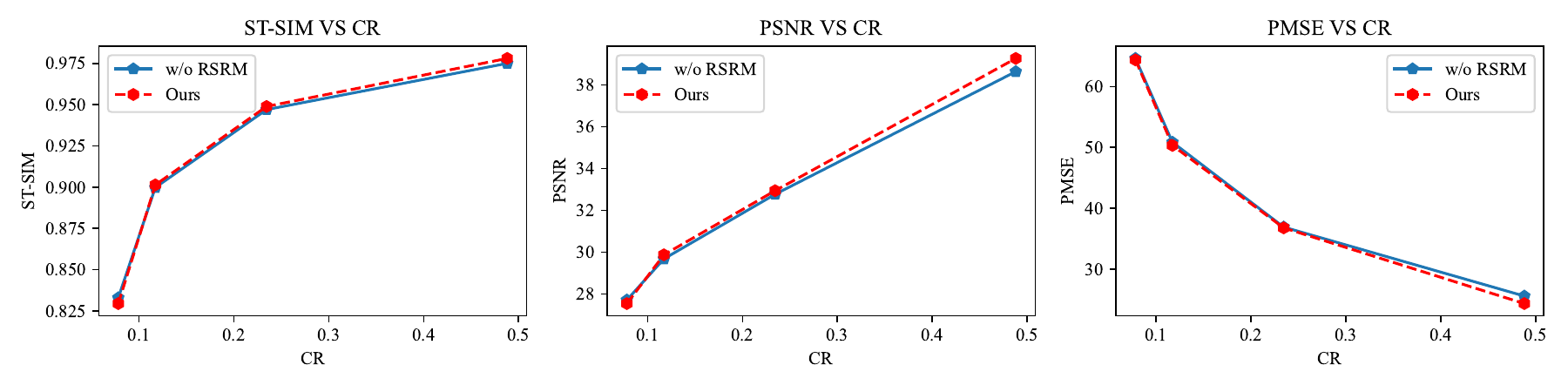}
    \caption{The performance of the model with and without RSRM under different CRs}
    \label{fig12}
\end{figure*}
\begin{figure*}[!t]
    \centering
    \includegraphics[width=1\linewidth]{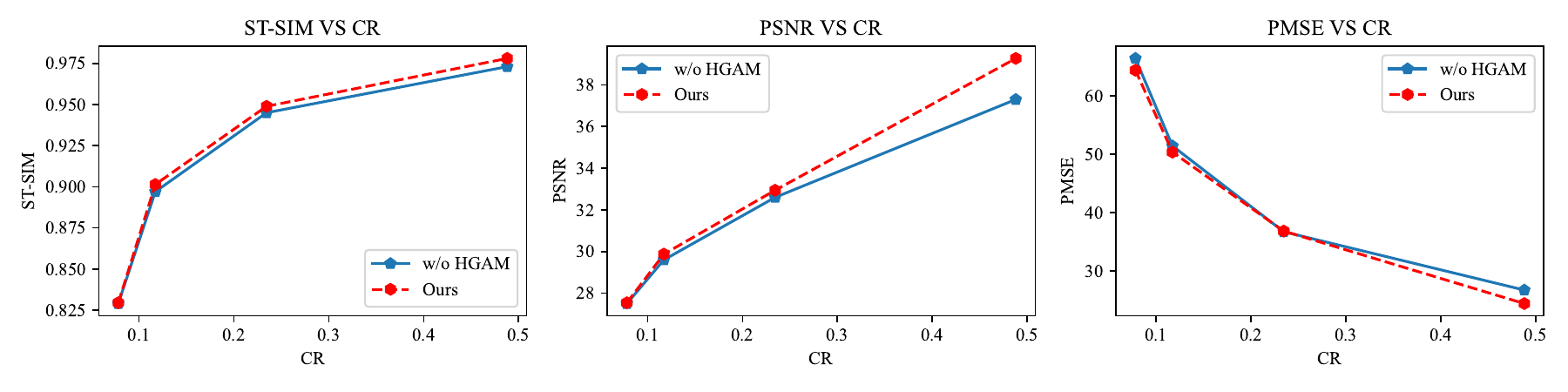}
    \caption{The performance of the model with and without HGAM under different CRs}
    \label{fig13}
\end{figure*}
To further evaluate the adaptability of our approach, we compare our model—capable of generating outputs across multiple compression ratios using a single set of model weights—with alternative models that necessitate separately trained weights for each specific compression ratio. As illustrated in Fig.~\ref{fig8}, regardless of whether the compression ratio is high or low, our model consistently outperforms DNN and DNN2 across all three evaluation metrics: ST-SIM, PSNR, and PMSE.

The results from the PSNR metric indicate that DNN and DNN2 do not adequately extract signal features, resulting in subpar reconstruction quality. In contrast, our model achieves optimal PSNR performance at low compression ratios, demonstrating its superior feature extraction capabilities and its ability to reconstruct vibrotactile signals with greater fidelity. Furthermore, this validates the model’s flexibility in rate adaptation and its robustness in signal representation.

Furthermore, we assessed the performance of our proposed model using the VCTK speech dataset. As illustrated in Fig.~\ref{fig9} and Fig.~\ref{fig10}, the results indicate the model's effectiveness across various channel SNR conditions and compression ratios. In comparison to other deep learning-based JSCC models for speech, our approach demonstrates competitive performance while maintaining relatively low model complexity, underscoring its practical applicability.
\subsection{Ablation Study}

To validate the effectiveness of the proposed model components, we conduct ablation studies using the IEEE 1918.1.1 standard vibrotactile dataset. Three key modules—CFPM, RSRM, and HGAM—are individually removed or replaced to evaluate their respective contributions. The resulting model variants are designated as w/o CFPM, w/o RSRM, and w/o HGAM.

As illustrated in Fig.~\ref{fig11}, the removal of CFPM leads to a significant decline in adaptability under low channel noise and a reduced robustness to SNR variations across all evaluation metrics, including ST-SIM, PSNR, and PMSE. This indicates the crucial role of CFPM in managing channel fluctuations. Replacing RSRM with a convolutional layer and batch normalization (w/o RSRM) yields comparable performance at high compression ratios; however, it leads to noticeable degradation at low compression ratios (Fig.~\ref{fig12}), suggesting RSRM effectively enhances feature extraction and reconstruction under challenging compression conditions. Furthermore, the removal of HGAM (w/o HGAM) consistently results in inferior performance across various compression levels (Fig.~\ref{fig13}), underscoring its importance in capturing vibrotactile features and improving overall fitting capability. Collectively, these results confirm that CFPM, RSRM, and HGAM each play a significant role in enhancing the robustness and performance of the proposed model.
\section{Conclusion}
\label{Conclusion}
We propose a framework for vibrotactile signals known as FD-JSCC. This model consists of two key components: the RSRM and the CFPM.
Experiments conducted on the IEEE 1918.1.1 standard vibrotactile dataset demonstrate that the proposed model exhibits exceptional performance and robustness across various channel noise conditions. It satisfies the stringent low-latency requirements of vibrotactile applications, dynamically adapts to fluctuating channel conditions, and supports flexible rate control. Users can modify the compression ratio according to the available bandwidth. At CR of $[2, 4, 8, 16]$—the model maintains high perceptual quality while utilizing only 38.9\% of the parameters employed by DeepSC-S. This efficiency makes it particularly suitable for deployment in resource-constrained haptic systems. Furthermore, the proposed model demonstrates strong performance when applied to speech datasets, indicating its robust generalization capabilities across different modalities.
\bibliographystyle{IEEEtran}
\bibliography{IEEEexample}
\end{document}